\documentclass{article}

\usepackage{amsmath}
\usepackage{amssymb}
\usepackage{amstext}
\usepackage{bigstrut}
\usepackage{bm}
\usepackage{booktabs}
\usepackage[bf,footnotesize]{caption}
\usepackage{chngpage}
\usepackage{color}
\usepackage{dcolumn}
\usepackage{float}
\usepackage{fullpage}
\usepackage{graphicx}
\usepackage{lineno}
\usepackage{makecell}
\usepackage{mathpazo}
\usepackage{mathrsfs}
\usepackage{mathtools}
\usepackage{multirow}
\usepackage{rotating}
\usepackage{setspace}
\usepackage{subfigure}
\usepackage{threeparttable}
\usepackage{xr}
\usepackage{algorithm}
\usepackage{algorithmic}

\definecolor{darkred}{rgb}{0.75,0,0}
\definecolor{darkgreen}{rgb}{0,0.5,0}
\definecolor{darkblue}{rgb}{0,0,0.75}
\definecolor{darkorange}{rgb}{1,0.9,0.1}
\definecolor{dark}{rgb}{0,0,0}

\usepackage{caption}
\newcommand*\patchAmsMathEnvironmentForLineno[1]{%
	\expandafter\let\csname old#1\expandafter\endcsname\csname #1\endcsname
	\expandafter\let\csname oldend#1\expandafter\endcsname\csname end#1\endcsname
	\renewenvironment{#1}%
	{\linenomath\csname old#1\endcsname}%
	{\csname oldend#1\endcsname\endlinenomath}}%
\newcommand*\patchBothAmsMathEnvironmentsForLineno[1]{%
	\patchAmsMathEnvironmentForLineno{#1}%
	\patchAmsMathEnvironmentForLineno{#1*}}%
\AtBeginDocument{%
	\patchBothAmsMathEnvironmentsForLineno{equation}%
	\patchBothAmsMathEnvironmentsForLineno{align}%
	\patchBothAmsMathEnvironmentsForLineno{flalign}%
	\patchBothAmsMathEnvironmentsForLineno{alignat}%
	\patchBothAmsMathEnvironmentsForLineno{gather}%
	\patchBothAmsMathEnvironmentsForLineno{multline}%
}

\makeatletter

\makeatletter

\begin{document}

\title{Strategy evolution on temporal hypergraphs}


\author{Xiaochen Wang$^{1,\dag}$, Lei Zhou$^{2, \dag}$, Alex McAvoy$^{3, 4}$, Zhenglong Tian$^{1}$, and Aming Li$^{1,5,\ast}$
\\
\footnotesize{$^{1}$Center for Systems and Control, College of Engineering, Peking University, Beijing 100871, China}\\
\footnotesize{$^{2}$School of Automation, Beijing Institute of Technology, Beijing 100081, China}\\
\footnotesize{$^{3}$School of Data Science and Society, University of North Carolina at Chapel Hill, Chapel Hill, NC 27599, USA}\\
\footnotesize{$^{4}$Department of Mathematics, University of North Carolina at Chapel Hill, Chapel Hill, NC 27599, USA}\\
\footnotesize{$^{5}$Center for Multi-Agent Research, Institute for Artificial Intelligence, Peking University, Beijing 100871, China}\\
\footnotesize{$\dag$ These authors contributed equally to this work}\\
\footnotesize{$\ast$ Corresponding author. E-mail: amingli@pku.edu.cn}\\
}

\maketitle


\begin{abstract}
Individuals interact and cooperate in structured systems. Many studies represent this structure using static networks, where each link represents a permanent connection between two nodes. However, real interactions are generally not time-invariant and are often not pairwise. Recently, progress has been made in modelling higher-order interactions using hypergraphs, where a link may connect more than two individuals. Here, we study cooperation through temporal hypergraphs, capturing the time-varying, higher-order interactions in empirical systems. We find that temporal hypergraphs can promote cooperation compared with static networks and the latter may underestimate the positive beneficial effects of constrained, local interactions on fostering cooperation. We further show that cooperation can be facilitated by temporal hypergraphs with sparse components and higher-order interactions. Importantly, when interactions match population size, relatively small hyperedge sizes best facilitate cooperation. Synthetic and empirical hypergraphs alike affirm our findings, illuminating how temporal, higher-order interactions profoundly shape the evolution of cooperation.
\end{abstract}

\section*{Introduction}

The prevalence of cooperation across taxa reflects a remarkable, albeit not universal, inclination toward altruism in many societies\cite{ESS,trivers1971evolution,axelrod1981evolution,nowak1992evolutionary,ohtsuki2006simple,santos2006evolutionary}.
Cooperation, however, which entails a personal cost and benefits others, is outperformed by defection (free-riders) in terms of both individual rationality\cite{nash1950equilibrium} and Darwin's notion of ``survival of the fittest''\cite{ hofbauer1998evolutionary} under natural selection, where the mechanisms for reciprocity are absent. It is well-known that the evolution of cooperation stems, in part, from limited dispersal\cite{hamilton1963evolution}--either in terms of genetic relatives or in terms of cultural adoption of behaviours. Network structure, a notable characteristic of human and non-human societies, for example as a typical dispersal mechanism, can constrain social interactions and the imitation of strategies\cite{santos2005scale,ohtsuki2006simple,allen2017evolutionary}.

In fact, network reciprocity is one of the fundamental explanations for the evolution of cooperation\cite{nowak2006five} since it enables cooperators to form clusters and avoid exploitation by defectors 
\cite{santos2006evolutionary,santos2008social,gomez2007dynamical,santos2005scale,nowak1992evolutionary,ohtsuki2006simple,nowak2006five,allen2017evolutionary,battiston2020networks,nowak2010evolutionary,fotouhi2019evolution}. Higher-order interactions, which occur among multiple individuals simultaneously, are prevalent in human and non-human systems, such as co-authorship in academia\cite{patania2017shape}, mesoscopic interactions among neurons in the nervous system\cite{ganmor2011sparse}, and competition among species in ecosystems\cite{levine2017beyond}.
Cooperative behavior in higher-order networks is also widely observed, such as women's cooperative foraging groups among hunter-gatherers\cite{jang2024women} and intergroup conflict among chimpanzees and other primates\cite{wilson2001does}.
Most previous studies about higher-order interactions on networks focus on classical network models, where nodes represent individuals and links represent interactions\cite{perc2013evolutionary} (Fig.~\ref{fig:eg}a).
A widely adopted model for capturing higher-order interactions involves each individual forming a group with all neighbours\cite{santos2008social,li2016evolutionary,perc2013evolutionary}.
Individuals interact with not only their neighbours but also their neighbours' neighbours.
However, while changing behaviours, individuals can only imitate their neighbours rather than all interaction partners, suggesting a limitation of pairwise networks in accurately describing higher-order interactions\cite{battiston2020networks,majhi2022dynamics}.
Recent work points out that hypergraphs (i.e., with hyperedges given by sets of nodes, generalizing the notion of ``link'' to more than two individuals) are more naturally suited to capturing higher-order interactions than their pairwise counterparts\cite{alvarez2021evolutionary,iacopini2019simplicial} (Fig.~\ref{fig:eg}b).
Indeed, it has been shown that the presence of higher-order interactions in hypergraphs has a fundamental impact on the evolution of cooperation\cite{alvarez2021evolutionary,civilini2021evolutionary,alvarez2022collective}.
Meanwhile, the nonlinearity of group games and the diversity of combinatorial structures also influence the evolution of cooperation at the level of higher-order interactions\cite{sheng2024strategy}.

Although hypergraphs provide a valid tool for studying the effects of higher-order interactions on the evolution of cooperation, current studies--like many on classical networks as well--often assume that these interactions exist permanently and never change.
In reality, interactions among individuals appear and dissolve over time, indicating transient, dynamic links.
For example, human interactions through digital or face-to-face communications usually last for finite, short durations\cite{toroczkai2007proximity}, and empirical ecological networks change with the seasons\cite{saavedra2016seasonal}.
More importantly, it has been shown that such network temporality can significantly alter various dynamical processes on networks, such as the spread of information\cite{masuda2013temporal}, the evolution of cooperation\cite{li2020evolution,su2023strategy}, and the controllability of systems\cite{li2017fundamental}.
A substantial body of research has highlighted the widespread presence of temporal and higher-order structures in population interactions across natural systems. For instance, higher-order interactions among competing species \cite{grilli2017higher} and seasonal variations in species interactions \cite{saavedra2016seasonal} provide strong evidence that temporal hypergraphs offer a structural model that more accurately reflects real-world populations. 
And there are extensive evidence that temporal higher-order structures are ubiquitous in real-world systems\cite{benson2018simplicial}, including biological, medical, social and web networks.
Simultaneously, research on other kinds of dynamics on temporal higher-order structures has already demonstrated profound impacts\cite{neuhauser2021consensus,iacopini2022group,majhi2022dynamics,burgio2020evolution,kumar2021evolution}, prompting us to investigate whether this novel insight will bring about fundamental changes to the study of the evolution of cooperation.

Here, we systematically investigate the evolution of cooperation on temporal hypergraphs, where hypergraph structures change over time.
By combining computational simulations and mathematical analysis on both synthetic and empirical temporal networks of human interactions, we find that temporal hypergraphs with higher-order interactions and sparse components can strongly promote cooperation, and even outperform their static counterparts.
Simultaneously, by comparing different modelling techniques on the same datasets, we validate that pairwise and static network models may underestimate the potential that network structures promote the emergence of cooperation.
Intriguingly, when we consider the limitations of order in human offline interactions and extend our focus to larger-scale higher-order interactions, we find that, in fact, interactions of moderately small order foster cooperation.
Larger hyperedges can actually be detrimental to cooperation.
Our results unveil intriguing facets of how temporal higher-order interactions, a realistic feature of many kinds of systems capable of exhibiting evolutionary dynamics, impact the evolution of strategies like cooperation.

\section*{Model}
\begin{figure}
\centering
\includegraphics[width=0.95\linewidth]{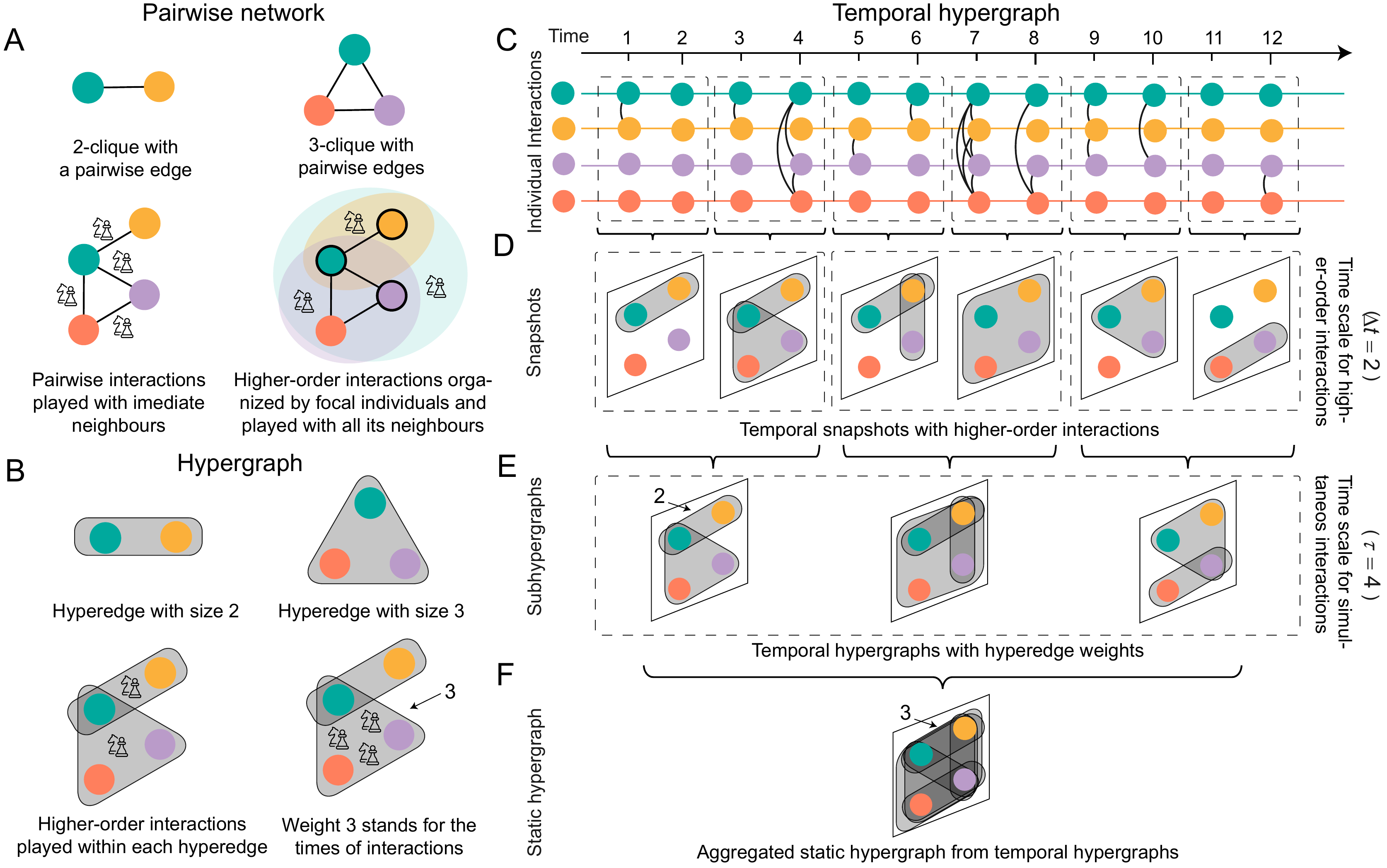}
\caption{\textbf{Construction of temporal hypergraphs.}
\textbf{a}, On classical networks, a link represents pairwise interaction and can connect only two individuals. 
For higher-order interactions on pairwise networks, each focal individual (black circled nodes) organizes a common pool in which everyone interacts once.
\textbf{b}, On hypergraphs, a hyperedge can connect more than two individuals.
Individuals within a hyperedge interact together, and the weight is used to indicate the number of interactions.
\textbf{c}, Social interactions among $4$ individuals are indicated by nodes with different colours.
Each individual is depicted by a line, over which the corresponding nodes will be connected at time $t$,
provided two players interact with each other during each time interval.
\textbf{d}, The snapshots are generated from all interactions during the time window, $\Delta t$.
Individuals are in a given hyperedge if they interact with each other during the window $\left(t-\Delta t,t\right]$.
\textbf{e}, The snapshots within each time interval, $\tau$, are then aggregated into subhypergraphs.
For example, the green and yellow nodes interact on the first and second snapshots, thus there is a hyperedge between them with weight $2$ on the first subhypergraph, which means they play the game twice.
\textbf{f}, All snapshots are aggregated into a static hypergraph.
}
\label{fig:eg}
\end{figure}

A hypergraph is a higher-order structure that consists of nodes and hyperedges. A hyperedge generalizes the notion of ``link'' in a graph, in that it can connect more than two individuals. To draw a clear distinction between the two, we refer to classical graphs as ``networks'' and higher-order graphs as ``hypergraphs.'' A temporal hypergraph is simply an ordered sequence of hypergraphs. As the hypergraph changes over time, hyperedges can appear or disappear (Fig.~\ref{fig:eg}).

Just as there are ample population structures in the classical setting, there are a great many ways in which temporal hypergraphs can arise. We first consider temporal hypergraphs arising from empirical datasets. The dataset can be modelled as a sequence of $\mathcal{G}$ snapshots, where each contains all interactions during non-overlapping time windows of length $\Delta t$, in a fixed population of $N$ individuals (Fig.~\ref{fig:eg}c, d). In each snapshot, every clique is represented by a hyperedge, in which individuals interact with each other (Fig.~\ref{fig:eg}d). The snapshots within each time interval, $\tau$, are then aggregated, generating a sequence of $\lceil\mathcal{G}/(\tau/\Delta t)\rceil$ subhypergraphs that form a temporal hypergraph (Fig.~\ref{fig:eg}e). When $\tau=\mathcal{G} \Delta t$, we obtain the corresponding static hypergraph (Fig.~\ref{fig:eg}f).

On temporal hypergraphs, individuals within the same hyperedge interact. We model the interactions as public goods games, where each cooperator ($C$) pays a cost, $c$, and each defector ($D$) pays nothing. The public goods game is a standard model of a multiplayer social dilemma, which are characterized by the properties that it always pays to switch to defection yet groups of cooperators are better off than groups of defectors\cite{dawes:ARP:1980}. In a standard public goods game, the total costs are then collected and multiplied by a synergy factor, $R>1$, to generate payoffs, which are evenly distributed to all individuals within the group. For a group of size $n$ with $n_{C}$ cooperators, the payoffs to defectors and cooperators are $Rcn_{C}/n$ and $Rcn_{C}/n-c$ respectively. Notably, by taking $r=R/n$, we then write these payoffs as $\pi_{D}\left(n_{C}\right) =rcn_{C}$ and $\pi_{C}\left(n_{C}\right) =\pi_{D}\left(n_{C}\right) -c$, which omit explicit division by the group size but are otherwise completely equivalent to the standard model of public goods games.

Extending public goods games to a population requires care because the size of the groups ($n$), which are represented by hyperedges, can vary throughout the population. As a result, the rescaling $R=rn$ leads to two natural options: either $R$ is independent of $n$ or $r$ is. If $R$ is fixed (and $r$ varies), then the strength of the social dilemma varies depending on the group size. 
A natural way to ensure social dilemmas are comparable across groups is to instead let $r=R/n$ be fixed. The latter model represents a good that is both non-excludable and non-rival, which forms the basis for the economic notion of a public good\cite{samuelson1954pure,rege2004social}. Relevant examples include public parks\cite{stewart2004city}, radio signals\cite{noam1995taking}, clean air\cite{levinson2012valuing}, and a stable climate\cite{tavoni2011inequality}. As a result, we focus our attention primarily on public goods games on hypergraphs such that $r=R/n$ is fixed as a function of group (hyperedge) size.

In each round, individuals accumulate net payoffs from their interactions. Individual $i$ simply adds up the payoffs obtained from all hyperedges $i$ belongs to in the hypergraph, yielding a net score of $u_{i}$. Next, individuals decide (synchronously) whether to change their strategies. Individual $i$ selects a random hyperedge and compares her payoff with that of a random neighbour, $j$, in that hyperedge. Here, individual $i$ imitates $j$'s strategy with probability given by the Fermi function, 
\begin{align}
e_{ji} &= \frac{1}{1+e^{-s\left(u_{j}-u_{i}\right)}} ,
\end{align}
where $s>0$ is the intensity of selection\cite{szabo1998evolutionary,traulsen2006stochastic}. Individuals play games with their neighbours for $g$ rounds on every subhypergraph before it changes to the next one. Note that individuals neither play games nor update their strategies, as they are not involved in any interactions when disconnected.

To determine whether the evolution of cooperation is promoted, we focus on the mean fraction of cooperators present in the population, $f_{C}$. As the size of the system is finite, the dynamics will eventually lead to an absorbing state (either full cooperation or defection). The mean fraction of cooperators $f_{C}$ represents the probability of the system reaching the state of full cooperation (see Box.~\ref{alg:pseudocode}). As the factor $r$ varies, we denote by $r_{C}$ the critical value such that $f_{C}>0$ whenever $r>r_{C}$. Smaller values of $r_{C}$ mean that cooperation more readily emerges in the population.

\section*{Results}

\subsection*{Higher-order interactions and sparse components}
\begin{figure}
\centering
\includegraphics[width=0.95\linewidth]{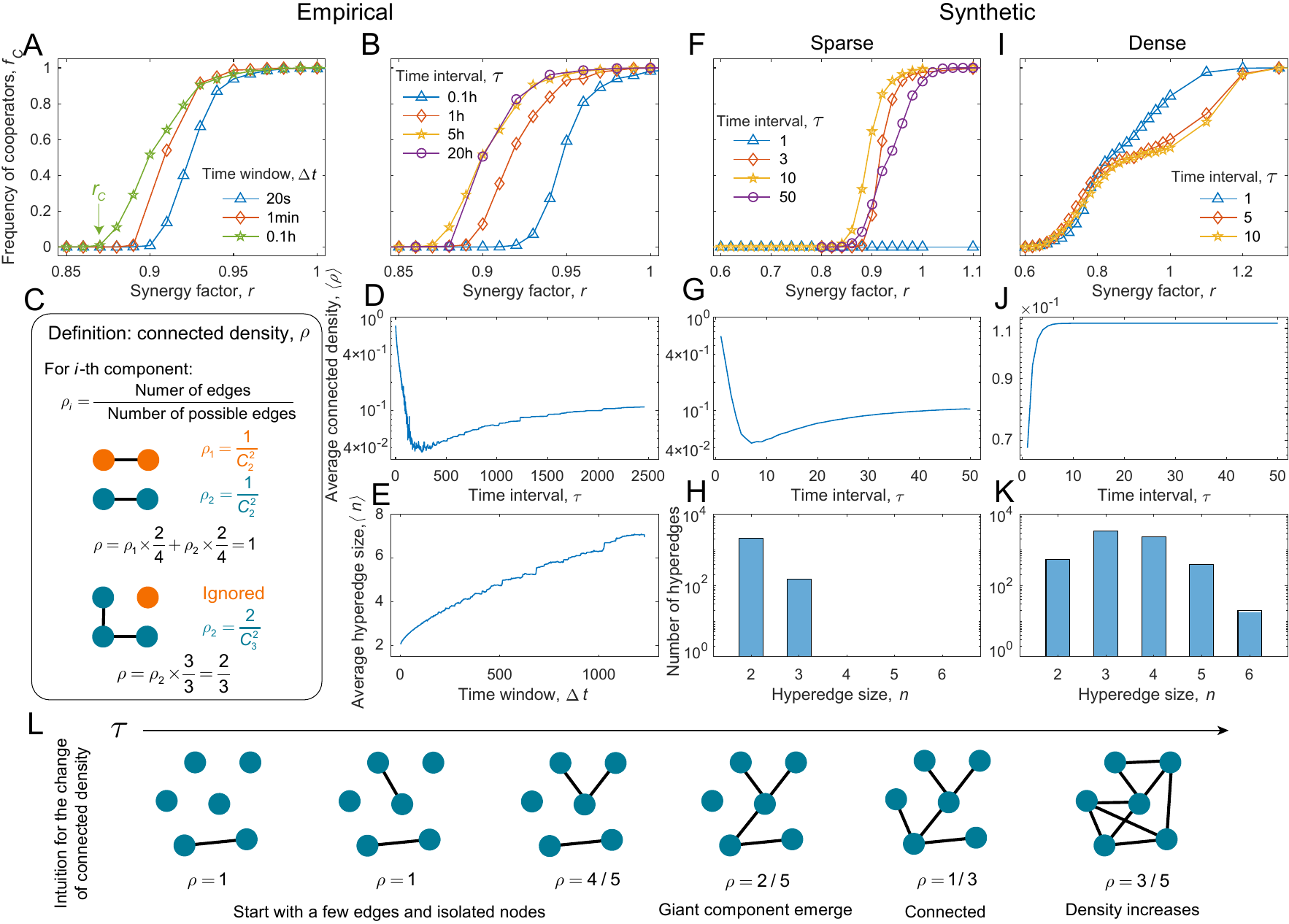}
\caption{\textbf{Higher-order interactions and sparse components together promote cooperation.}
    We show the fraction of cooperators, $f_C$, as a function of the synergy factor, $r$, on empirical (\textbf{a}, \textbf{b}) and synthetic (\textbf{f},\textbf{i}) temporal hypergraphs.
    For synthetic hypergraphs, we set the probability for links to be activated, $p$, to be $0.05$ (\textbf{f}) and $0.6$ (\textbf{i}).
    Cooperation emerges at a smaller value of $r_C$ when $\Delta t$ is large and $\tau$ is moderate.
    \textbf{c} illustrates the definition of connected density.
    For these two networks, with $4$ nodes and $2$ links, the traditional density is $1/3$.
    However, connected density focuses on the connected components.
    In the top network, there are two components with densities equal to $1$. The connected density of the whole network is also $1$.
    The other network (bottom), with one component and one isolated node, has a connected density of $2/3$.
    We present the average connected density, $\left\langle \rho \right\rangle$, as a function of time interval, $\tau$, on empirical (\textbf{d}) and synthetic (\textbf{g},\textbf{j}) hypergraphs.
    The unit of $\tau$ is $1$min in panel \textbf{d}.
    For the empirical graphs and the synthetic hypergraphs with $p=0.05$, the connected density drops sharply and then increases slowly as $\tau$ increases.
    However, it increases monotonically on the synthetic hypergraphs when $p=0.6$.
    \textbf{e}, We plot the average hyperedge size, $\left\langle n \right\rangle$, as a function of the time window, $\Delta t$, for the empirical hypergraphs.
    Notably, this size increases as the time window gets longer.
    We also display distributions of the hyperedge size for synthetic hypergraphs (\textbf{h}, \textbf{k}).
    The synthetic hypergraphs with $p=0.6$ have larger hyperedge sizes compared to those with $p=0.05$.
    \textbf{l},
    We provide intuition for the change of connected density as a function of $\tau$.
    The temporal network starts with isolated nodes and only a few links. As $\tau$ increases, a giant component emerges, and then the network becomes connected.
    Beyond this point, increasing $\tau$ can only increase the density.
    The last subhypergraph is ignored if $\tau$ does not divide $G\Delta t$.
    The simulations are performed $10^3$ times, and within each simulation, the evolutionary process runs for $10^6$ rounds.
    All simulations start with an equal number of cooperators and defectors, with the round of interactions $g=100$.
    The population size $N=327$ for the empirical hypergraphs and $N=100$ for the synthetic hypergraphs.
    We set $\tau=5$h for panel \textbf{a} and $\Delta t=0.1$h for panel \textbf{b}.
    The selection intensity is $s=2$ across all simulations.
}
\label{fig:tem_den}
\end{figure}

We first focus on the impact of population structure on the evolution of cooperation, such as density and group size. Figure~\ref{fig:tem_den}a shows that when the time interval $\tau$ of the subhypergraphs is constant, an increase in the time window $\Delta t$ both promotes the frequency of cooperators, $f_{C}$, and decreases the critical value, $r_{C}$.
For fixed $\Delta t$, as shown in Fig.~\ref{fig:tem_den}b, cooperation is facilitated when a subhypergraph contains interactions within a moderate interval (i.e., moderate $\tau$).
Intuitively, the window $\Delta t$ and interval $\tau$ shape the structural characteristics of the temporal sequence of hypergraphs.
Enlarging the window, $\Delta t$, allows more interactions to be included in the snapshots, making the sizes of hyperedges larger and the degree (i.e., the number of hyperedges each individual belongs to) larger.
Large $\tau$ can increase the degree and size of components on subhypergraphs, which, in turn, affects the aggregation of cooperators.

To intuitively explain the impact of temporal hypergraphs on the evolution of cooperation, we begin with an analogy involving a degree-homogeneous uniform temporal hypergraph, in which all individuals have the same degree and all hyperedges have the same size, $n$. When transitioning from one hypergraph to the next, the hyperedges remain unchanged with probability $\delta$ and are randomly reshuffled with probability $1-\delta$. Thus, $\delta$ quantifies the similarity between consecutive hypergraphs. Using a mean-field model, the frequency of cooperators, $x_{C}$, satisfies the equation
\begin{align}
\dot{x}_{C} &= x_{C}\left(\left(1-\delta\right)\left(1-x_{C}\right) +\delta q\right)\tanh\left(\frac{1}{2}s\left(\pi_{C}-\pi_{D}\right)\right) , \label{eq:replicator}
\end{align}
where $q$ is a parameter related to the probability that neighbours are using different strategies in the unchanged hyperedges (see Supplementary Information S1 for details). Here, $q$, $\pi_{C}$, and $\pi_{D}$ all depend on $x_{C}$. This is a version of the classical replicator equation\cite{taylor:MB:1978}, which can provide useful insights into the process (even if just a rough approximation of the actual dynamics).

One feature of this equation is that the equilibrium $x_{C}=0$ is stable for sufficiently small $r$ and unstable for sufficiently large $r$. In other words, there exists a critical multiplication factor,
\begin{equation}
r_{C} = \frac{1}{1+\delta\left(n-1\right)\left(1-q\right)} ,
\label{critical_r}
\end{equation}
such that the dynamics support a non-zero fraction of cooperators for $r>r_{C}$.

However, from Eq.~(\ref{critical_r}), one cannot immediately deduce that increasing $n$ always leads to a lower $r_{C}$ because $q$ itself depends on $n$ in a complicated fashion. Previous studies indicate that sparser networks (lower network density) lead to a more pronounced aggregation phenomenon, representing smaller values of $q$ in connected static networks\cite{ohtsuki2006replicator,tarnita2009strategy,gomez2007dynamical,gomez2007paths}. As such, Eq.~(\ref{critical_r}) provides intuition about the effects of network density on the evolution of cooperation in public goods games, using $q$ as a proxy, but a more direct approach would be to study the effects of network density directly, which we turn to next.

To better represent connectivity among the nodes, we use the concept of network density. First, we map the hypergraph to an unweighted network (i.e., projected network generated via expanding hyperedges into complete cliques).
For example, the hypergraph in Fig.~\ref{fig:eg}b can be projected as the network in Fig.~\ref{fig:eg}a.
The traditional definition of network density is the ratio between the number of existing links, $L$, and the number of possible links, $\binom{N}{2}$.
In a temporal hypergraph, the subhypergraphs may not be connected, and this metric may no longer be applicable.
Therefore, we use the concept of connected density of networks to capture the internal structure better.
For the $i$th component with $N_i$ nodes and $L_i$ links, the density is $\rho_{i}=L_{i}/\binom{N_{i}}{2}$, which lies between $0$ and $1$.
Note that we ignore the isolated nodes, as neither interaction nor dynamics occurs.
Then, the connected density of this subgraph is defined as the corresponding weighted average over each component, $\rho =\sum_{i=1}^{N}\left(N_{i}/N\right)\rho_{i}$.
For the temporal hypergraph, the average density, $\left\langle\rho\right\rangle$, is obtained by averaging $\rho$ over the relevant sequence of subhypergraphs.

This definition of density focuses on connectivity within components of a subhypergraph, which constrains where behaviours can spread. Networks that are extremely sparse under the traditional definition can still be dense from the perspective of their connected density. For example, if a network is composed of just pairwise components consisting of two nodes each, the traditional density will be low (sparse), but the connected density is equal to 1 (dense); see Fig.~\ref{fig:tem_den}c. The importance of this distinction is that, in such networks, the payoffs of cooperators are consistently lower than those of defectors, making the evolution of cooperation seemingly impossible.

We next explore the connected density and hyperedge size induced by $\tau$ and $\Delta t$.
As the time interval $\tau$ of the subhypergraph gets longer, the average connect density of the temporal hypergraph decreases rapidly and then gradually increases.
The connected density reaches a minimum at an intermediate value of $\tau$ (Fig.~\ref{fig:tem_den}d).
That is, for a moderate $\tau$, the hypergraph is sparser, which can facilitate the aggregation of cooperators.
If $\tau$ is extremely small, an increase in $\tau$ significantly increases the size of the components compared to the degree of connectivity between nodes within the components.
However, once the entire network is close to being fully connected, increasing $\tau$ only enlarges the number of links, thereby increasing the connected density.
Meanwhile, increasing $\Delta t$ leads to an increase in hyperedge size (Fig.~\ref{fig:tem_den}e), and Eq.~(\ref{critical_r}) shows that increasing the size of hyperedges tends to reduce the critical value, $r_{C}$, which is consistent with both our simulations (Fig.~\ref{fig:tem_den}a, b) and previous findings\cite{alvarez2021evolutionary}.
Besides, we analyze two other empirical datasets and conduct further analysis on the parameters across all datasets (Figs.~S1-S8).
The results are consistent with those mentioned above.

We perform additional simulations on synthetic temporal hypergraphs, constructed by activating links independently with probability $p$ (see Methods).
The temporal hypergraph with $p=0.05$ is disconnected in most snapshots and has similar properties and performance to the empirical hypergraph.
Cooperation is promoted at moderate $\tau$ (i.e., sparser hypergraphs), which coincides with the trend that the connected density first decreases rapidly and then begins to rise as $\tau$ increases (Figs.~\ref{fig:tem_den}f, g).
The temporal hypergraph with $p=0.6$ verifies our conclusions from the other side of the spectrum, in which all snapshots are fully connected.
As $\tau$ increases, the subhypergraphs become denser (Fig.~\ref{fig:tem_den}j) and inhibit cooperation (Fig.~\ref{fig:tem_den}i).
However, the size of hyperedges is larger for $p=0.6$ (Fig.~\ref{fig:tem_den}k), making it easier for cooperation to emerge on temporal hypergraphs (Figs.~\ref{fig:tem_den}i).
These findings further validate the influence of connected density, induced by temporal characteristics of the hypergraphs, on the evolution of cooperation.

\subsection*{Temporal versus static hypergraphs}
Having found that a combination of higher-order interactions and low connected density can promote cooperation, we now turn to the impacts of temporality on the evolution of cooperation. In relation to static hypergraphs in terms of model expressiveness, temporal hypergraphs better capture empirical interactions by incorporating the interaction time\cite{li2020evolution}.
Moreover, the existence of a temporal scale implies that the speed of interactions and the rate of structural changes can differ.

\begin{figure}
\centering
\includegraphics[width=0.65\linewidth]{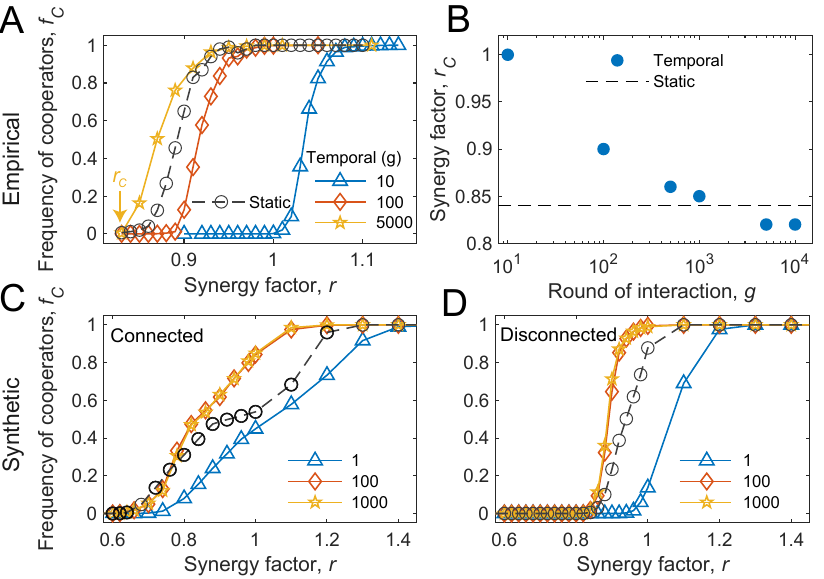}
\caption{\textbf{Temporal hypergraphs may promote cooperation compared to their static counterparts.}
We plot the fraction of cooperation, $f_C$, as a function of the synergy factor, $r$, on empirical (\textbf{a}) and synthetic (\textbf{c}, \textbf{d}) hypergraphs.
For synthetic hypergraphs, we set the probability $p$ for links to be activated to be $0.6$ (connected, \textbf{c}) and $0.05$ (disconnected, \textbf{d}).
The fraction of cooperators is promoted as the round of interactions $g$ increases and eventually exceeds that of the static counterpart.
\textbf{b},
We present the critical value of the synergy factor as a function of $g$. 
The magnitude of the promotion of cooperation (i.e., the reduction in $r_C$) diminishes as $g$ increases.
For panel \textbf{a}, $\Delta t=0.1$h, $\tau=1$h. In panels \textbf{c} and \textbf{d}, $\tau=1$ and $\tau=10$, respectively.
Other parameters are consistent with those of Fig.~\ref{fig:tem_den}.
}
\label{fig:g}
\end{figure}

Let $g$ be the number of time steps a subhypergraph persists before a transition. In other words, each subhypergraph (Fig.~\ref{fig:eg}e) is used for $g$ consecutive steps, with each involving payoff accumulation and then imitation of strategies. We find that enlarging the interaction time, $g$, facilitates cooperation, even transcending the effects of static counterparts (illustrated in Fig.~\ref{fig:eg}f). 
We have already seen that, for a proper time interval, $\tau$, the connected density is lower than that of the static hypergraph, which provides an avenue for the temporal hypergraph to promote cooperation.
The parameter $g$ is related to the similarity, $\delta$, appearing in the replicator equation (Eq.~(\ref{eq:replicator})).
Larger $g$ corresponds to $\delta$ closer to $1$.
Thus, from Eq.~(\ref{critical_r}), we anticipate that increasing $g$ will enhance cooperation.
Numerical results confirm this prediction (Fig.~\ref{fig:g}a),
and we find that, for a series of temporal hypergraphs generated by choosing an appropriate $\tau$ (corresponding to sparser subhypergraphs), temporal hypergraphs facilitate cooperation more than their denser static counterparts when $g$ is large.
As Fig.~\ref{fig:g}a shows, for $\Delta t=0.1$h and $\tau=1$h, such a transition occurs between $g=100$ and $g=5000$. 
Note that the results from the other two datasets are also consistent with those mentioned above (Fig.~S9).
Meanwhile, such results hold across networks of different sizes (Fig.~S11).

We further show that the magnitude of the promotion decreases with more rounds of interaction, $g$ (Fig.~\ref{fig:g}b).
As $g$ increases, cooperation is greatly facilitated, and the critical value $r_{C}$ drops.
But when $g$ continues to increase, this effect slows.
This observation can be explained by the fact that as $g$ increases, the rise of similarity $\delta$ of the temporal hypergraphs gradually slows down.
This effect is approximately proportional to $1/g$, as it is equivalent to switching to a new one after experiencing $g$ identical subhypergraphs (Fig.~S10).
The same situation is observed in the synthetic temporal hypergraphs (Figs.~\ref{fig:g}c, d).
When $g=100$, all temporal hypergraphs outperform their static counterparts.
But by further increasing $g$ from $100$ to $1000$, the additional gain in the reduction of $r_{C}$ is negligible.

Intuitively, once the structure is defined, a fast evolutionary process, which is equivalent to having a long and fixed period of interaction (i.e.~large $g$), helps the spread of cooperation.
Such a phenomenon is directly linked to the promotion of aggregations. 
When $g$ is larger, there is more time for cooperators to cluster. On the other hand, rapid changes in subhypergraphs (i.e., small $g$) can disrupt newly formed aggregations, thus enabling exploitation and inhibiting the spread of cooperation.
However, our findings also indicate that $g$ does not need to be particularly large.
A moderate $g$ can already allow for the advantageous aggregation of cooperators to be fully realized.

\subsection*{Advantages of higher-order networks}
We now contrast the effects of higher-order graphs with traditional pairwise networks.
Specifically, we compare higher-order interactions on hypergraphs with classical pairwise interactions on traditional networks.	
Following previous work\cite{santos2005scale,santos2006evolutionary,li2020evolution}, we construct traditional temporal networks under the same window $\Delta t=0.1$h and time interval $\tau=1$h through empirical data sets, where the cliques are no longer modelled as hyperedges\cite{li2020evolution}.
The game is the same as that on hypergraphs (i.e., public goods), but on networks, the participants in each game are two individuals sharing a link.

Our findings indicate that representing group interactions as a set of pairwise interactions may lead to an underestimation of the potential for local interactions to promote cooperation.
We find that hypergraphs may significantly facilitate cooperation compared to traditional networks based on pairwise interactions for different values of $g$ (Fig.~4). 
Cooperation emerges with a smaller synergy factor $r_C$ in both temporal and static hypergraphs.
The same results are also observed in the other two datasets (Fig.~S12).
In terms of Eq.~(\ref{critical_r}), pairwise links on traditional networks are smaller in size (smaller $n$), while at the same time, each individual has a larger number of neighbours (larger $q$) compared with hypergraphs. Both factors, we find, may inhibit the evolution of cooperation. Through a comparison of these four classes of networks, we confirm the facilitating role of higher-order interactions on the evolution of cooperation.

\subsection*{Evolution of cooperation with macro-interactions}
Our findings so far indicate that large hyperedge sizes tend to promote cooperation.
An intriguing question then arises: should we continuously increase the scale of group interactions?
If so, the evolution of cooperation would be maximally facilitated when all individuals are contained within a single hyperedge.
However, we note that the previous scenarios involve an assumption, due to natural limitations of human interaction datasets as well as the mean-field analysis leading to Eq.~(\ref{eq:replicator}), that the population size is significantly larger than the hyperedge size, i.e., $N\gg n$ (Figs.~\ref{fig:tem_den}e, h, k).
Hence, we next consider macro-interactions, where the hyperedge size is relatively large compared to the population size.
This scenario is also common in empirical observations.
For example, members with diverse research directions, interests, or grades (higher-order interactions) within the same research group (population), often benefit from gathering together for collective discussions. 
In such cases, the size of higher-order interactions relative to the population size can no longer be considered negligible.

To investigate this scenario, we design a synthetic hypergraph. 
From a pool of $N$ individuals, we randomly select $n$ individuals to form a hyperedge and repeat this process a total of $m$ times to create one subhypergraph. 
We then follow the same procedure to construct temporal hypergraphs with a total of $\mathcal{G}$ subhypergraphs under different hyperedge size $n$.
We conduct simulations on the resulting temporal sequence of hypergraphs for different values of $n$ (Fig.~\ref{fig: 5}a).
The results show that as the size of hyperedges increases, cooperation is initially enhanced and subsequently suppressed. 
While a larger hyperedge size can indeed reduce the critical value of $r_C$ required for cooperation to emerge (Fig.~\ref{fig: 5}b), the multiplication factor at which cooperation emerges is not the sole metric.
The rate of the fraction of cooperators, $f_C$, goes up with the increase in the synergy factor, $r$, is equally important.
For example, we observe that cooperation can initially emerge with lower $r$ when hyperedge sizes are larger, but as $r$ increases, $f_{C}$ grows larger in the population with smaller hyperedge.
To quantify this effect, we introduce another critical point, $r_{99}$, which is the synergy factor for which $f_{C}$ reaches $0.99$. 
We find that increasing $n$ only slightly reduces $r_{C}$ (Fig.~\ref{fig: 5}b), but significantly reduces the slope (Fig.~\ref{fig: 5}c) and increases $r_{99}$ (Fig.~\ref{fig: 5}d), thus making it more challenging for the population to achieve a state of almost full cooperation.
This is different from the case when $N \gg n$, where the relationship between the numerical values of $r_{C}$ and $r_{99}$ are essentially consistent across different parameters.
Moreover, as the population size $N$ increases, the value of hyperedge sizes $n$ that most effectively promotes cooperation may vary  (Fig.~S18). 
However, in general, relatively small values of $n$ tend to promote cooperation more effectively.
Therefore, moderately small hyperedges are most effective in promoting widespread cooperation.

We can understand this finding more intuitively by noting that, when $N\gg n$, the density of the hypergraph, hyperdegree, and the size of hyperedges are approximately independent. However, when dealing with macro-interactions, an increase in the size of hyperedges inevitably leads to an increase in density, as it implies that all $n$ individuals are connected with one another.
Simultaneously, since each individual may be part of more than one hyperedge, the average hyperdegree of the graph also increases.
All these properties can impede the evolution of cooperation.

Finally, we validate these conclusions on empirical temporal hypergraphs. 
We utilized an intergovernmental organizations (IGO) dataset\cite{pevehouse2020tracking}, comprising international organizations involving 193 countries and regions.
We transform this dataset into a static hypergraph (countries and regions for nodes, organizations for hyperedges, shown in Fig.~\ref{fig: 6}a) and, following the same methodology used for constructing the synthetic temporal hypergraph, convert it into a temporal hypergraph (see Methods).
Specifically, we categorize hyperedges into two types: small ($n<40$) and large ($n\ge 40$).
In each subhypergraph, the activation probability for large hyperedges is denoted by $p_{\text{L}}$, while for small hyperedges, it is represented by $p_{\text{S}}$.
These parameters can be viewed as the probabilities of interaction (e.g., organizations convening meetings) between hyperedges of different sizes.

We perform simulations on temporal hypergraphs with different values of $p_{\text{L}}$ and $p_{\text{S}}$.
If the probability for small hyperedges to be activated is fixed, making large hyperedges more active significantly inhibits cooperation (Fig.~\ref{fig: 6}b).
This is because the presence of large hyperedges notably increases the hypergraphs' connected densities, making it easier for free-riders to take advantage.
Conversely, if the activation probability of large hyperedges remains constant, a higher number of active small hyperedges will enhance cooperation (Fig.~\ref{fig: 6}c).
It is worth noting that when small hyperedges are not activated, cooperation remains at a very low level.
In this scenario, small hyperedges reinforce the connections between certain nodes, representing an increase in the weight of connections between these points (which is overlooked in the definition of connected density due to its complexity). 
High-weighted connections can facilitate reciprocity among nodes and, again, foster the aggregation of cooperators, which aligns with previous findings on static networks\cite{allen2017evolutionary}.
The above conclusions also apply to different values of $p_L$ and $p_S$ (Figs.~S14, S15).

\section*{Discussion}
%

Group (higher-order) interactions are naturally more complicated than pairwise encounters\cite{friedman2017community}, and the outcomes of multiplayer games cannot be fully predicted by pairwise models\cite{gokhale2010evolutionary}.
Our research indicates that the effects of high-order interactions exhibit complexity and nonlinearity.
Note that the proposed framework of evolutionary dynamics on temporal hypergraphs not only applies to public goods games but also to multi-strategy\cite{tarnita2011multiple,gokhale2014evolutionary,sinervo1996rock}, multi-player games with nonlinear and general payoff functions\cite{zhou2015evolution,li2016evolutionary}, which are important models of frequency-dependent behaviours.

For human empirical interactions, the temporality of higher-order interactions is encoded in three timescales: $\Delta t$, $\tau$, and $g$. $\Delta t$ represents the temporal scale of higher-order interactions, indicating how long pairwise interactions can be considered as defining higher-order interactions. 
Considering the characteristics of human interactions and the empirical data we use, this timescale is normally short. 
An increase of $\Delta t$ can enlarge the size of higher-order interactions and thus promote cooperation.
The parameter $\tau$ denotes the timescale of the subhypergraphs, signifying how long interactions within a timeframe can be regarded as occurring simultaneously. 
A higher ratio of $\tau$ to $\Delta t$ implies a large time span for the subhypergraph, which alters its density. 
An intermediate $\tau$ signifies a moderate level of simultaneous interactions, which can facilitate cooperation. 
The timescale parameter $g$ is the ratio between the interaction frequency and switching frequency of the subhypergraphs. 
A larger $g$ represents slower structural change (or more frequent interactions), while a smaller value implies rapid structural changes. 
Our findings indicate that slow structural change can promote the formation of cooperator aggregations, thereby fostering cooperation.
Notably, our conclusions also hold across different game types and evolutionary dynamics (see Supplementary Information S4).
The results are also robust to real world noises in datasets (see Supplementary Information S3.5).

Our results indicate that, in fact, a moderate hyperedge size is best for cooperation. 
In Supplementary Information S9, we confirm this effect on static uniform ring hypergraphs (Figs.~S16-S18), which ensures that every node has a hyperdegree of $2$.
For real-world human interactions, it may often hold that $N \gg n$, as space and time are limited, so increasing the interaction size encourages cooperation. 
However, if macro-interactions exist, larger sizes can strongly impede cooperation, as this allows a defector to exploit more cooperators. 
With the development of the internet, human online interactions can indeed involve large hyperedges\cite{ellison2007benefits,shirky2008here}. 
Our findings may also generate a possible explanation for why human interactions are often more amicable offline, while online interactions can easily manifest hostile behaviours\cite{shachaf2010beyond,tufekci2017twitter}.
Another example is international cooperation.
Our results indicate that smaller multilateral organizations may be more effective in promoting cooperation compared to larger ones. 
Nevertheless, in certain games, such as climate change and preventing the proliferation of nuclear weapons, all countries are compelled to participate. 
Our findings suggest that if the presence of macro-interactions (organizations) cannot be avoided, establishing more--and smaller--organizations can promote cooperation because the national interests within them are more closely aligned.
A direction worth exploring is asymmetric interactions, which are particularly pronounced in contexts like international organizations, where the costs individuals can incur and the distribution of benefits vary\cite{hauser2019social,chen2024cooperation}. Such a system may have profound implications for the evolution of cooperation.

Previous studies have demonstrated that both temporal structural variations and higher-order interactions can promote cooperation\cite{li2020evolution,alvarez2021evolutionary}. However, the synergistic effect between the two, as depicted in Fig.~4, is nonlinear. 
In the scenario of Fig.~4a, both higher-order interactions and temporal changes enhance cooperation compared to static pairwise interactions. 
However, introducing temporal dynamics into higher-order interactions inhibits cooperation, with the temporal hypergraph promoting cooperation only when $g$ is sufficiently large.
Our work reveals the interplay between higher-order interactions and temporal graphs and their impact on the evolution of cooperation.

\section*{Methods}
\subsection*{Degree-homogeneous uniform temporal hypergraphs}
To provide an intuitive analogy, we consider a degree-homogeneous uniform temporal hypergraph. 
In this kind of temporal hypergraph, each node has the same degree, and the size of each hyperedge is also uniform.
Let $\delta$ represent the average similarity between consecutive subhypergraphs. 
In this sequential process, a fraction, $1-\delta$, of the hyperedges in the last subhypergraph undergoes random reorganization to create a new subhypergraph. Within the newly generated hyperedges, the probability of an individual's neighbour being a cooperator is $x_{C}$, while the probability of them being a defector is $1-x_{C}$. The remaining hyperedges are unchanged, facilitating the aggregation of individuals with the same strategies, and increasing the likelihood of individuals and their neighbours sharing the same strategy. Further details may be found in Supplementary Information.

\subsection*{Constructing hypergraphs from empirical datasets}
For human interactions, we consider three empirical datasets\cite{fournet2014contact,mastrandrea2015contact,genois2015data}. Each dataset represents active contacts among a fixed population of $N$ individuals in intervals within $20$ seconds. 
These interactions are grouped into snapshots over a time window, $\Delta t$, which is a multiple of $20$ seconds, resulting in a sequence of $\mathcal{G}$ snapshots. 
Snapshots without interactions are excluded. 
Hyperedges are created based on these snapshots, where each clique represents a hyperedge representing interactions within the specified time window, $\Delta t$. 
These snapshots are further aggregated, with $\tau /\Delta t$ snapshots forming one subhypergraph. 
The total number of subhypergraphs is $\left\lceil\mathcal{G}/\left(\tau /\Delta t\right)\right\rceil$, and it is worth noting that if two snapshots within time $\tau$ share the same group of individuals in their hyperedges, they are considered as two separate hyperedges with a weight of $2$. 
Thus, they interact two times, and the probability of them choosing this hyperedge also increases during strategy updating.
Therefore, the temporal hypergraph is a sequence of subhypergraphs, which change from step to step.

For datasets pertaining to intergovernmental organizations (IGOs), the original dataset comprises a compilation of members associated with various IGOs\cite{pevehouse2020tracking}. 
In our analysis, we have specifically chosen to focus on a subset of this data, encompassing $193$ distinct countries and regions, as well as $309$ different IGOs.
We have depicted the distribution of sizes of IGOs in Supplementary Fig.~S19.
We treat each country as a distinct node, while each IGO is considered as a hyperedge connecting these nodes. 
In this context, if a country is a member of a particular IGO, we treat the corresponding node as part of the hyperedge.
In order to model the temporal dimension of our analysis, we denote the number of subhypergraphs as $\mathcal{G}$, which we set to $12$ to emulate the $12$ months in a calendar year. 
Within each subhypergraph corresponding to a month, a hyperedge is subject to activation with a certain probability.
We categorize the hyperedges into two distinct types: ``large'' hyperedges, defined as those with a membership size ($n$) greater than or equal to $40$, and ``small'' hyperedges of size less than $40$. 
Each category of hyperedges is associated with a unique activation probability, denoted $p_{\text{L}}$ for large hyperedges and $p_{\text{S}}$ for small hyperedges.

\subsection*{Constructing synthetic hypergraphs}
In order to construct networks exhibiting typical characteristics of empirical networks, such as high clustering coefficients\cite{watts1998collective} and scale-free degree distributions\cite{barabasi1999emergence}, we apply the Holme-Kim model\cite{holme2002growing}, which generates a static network.
To construct each snapshot for the synthetic temporal hypergraph, the links of the static network are activated with a probability $p$,
and all the resulting cliques are represented by hyperedges.
This process is repeated $\mathcal{G}$ times to construct each snapshot.
Subsequently, a series of synthetic subhypergraphs are constructed and form the synthetic temporal hypergraph through the procedure described previously for empirical temporal hypergraphs.

\subsection*{Simulation process}
The simulation process is shown in Box.~\ref{alg:pseudocode}.
At the beginning of the $k$th replicate, half of the nodes in the population are randomly designated as cooperators. 
In each round, every node engages in a game within all its hyperedges and accumulates corresponding payoffs. 
After interactions, all individuals simultaneously update their strategies.
Each node randomly selects one of its hyperedges, then randomly selects a neighbour within this hyperedge for a payoff comparison and (potential) strategy imitation.
After all nodes have updated their strategies, the number of cooperators is counted to obtain the cooperator frequency $f_{C,k}^{t}$ for the current round, $t$. 
If cooperators dominate ($f_{C,k}^{t}=1$) or disappear ($f_{C,k}^{t}=0$) from the population, the simulation ends. 
Otherwise, after $4 \times 10^6$ rounds, the average cooperator frequency over the last $1000$ rounds is recorded as $f_{C,k}$. 
The process is repeated $1000$ times, and the average cooperator frequency is recorded as $f_{C}$.

\begin{algorithm}[!h]
\caption{Pseudocode for simulating the model.}
\label{alg:pseudocode}
\begin{algorithmic}[1]
\FOR{Synergy factor $r$}
    \STATE \textbf{Start Simulation}
    \FOR{Each replicate $k=1\dots 1000$}
        \STATE Randomly select half of the nodes in the population as cooperators.
        \FOR{Each round $t=1\dots 4 \times 10^6$}
            \FOR{Each node $i=1\dots N$}
                \FOR{Each hyperedge containing $i$}
                    \STATE Accumulate payoffs: $u_{i}\gets u_{i}+rcn_C$ if $i$ is a defector and $u_{i}\gets u_{i}+rcn_C-c$ if $i$ is a cooperator, where $n_C$ is the total number of cooperators in the hyperedge.
                \ENDFOR
            \ENDFOR
            \FOR{Each node $i=1\dots N$}
                \STATE Randomly select one of its hyperedges.
                \STATE Randomly select a neighbour within this hyperedge, $j$.
                \STATE Compare payoffs and imitate $j$'s strategy with probability $1/\left(1+e^{-s\left(u_{j}-u_{i}\right)}\right)$.
            \ENDFOR
            \STATE Count the number of current cooperators to obtain the cooperator frequency $f_{C,k}^{t}$.
            \IF{$f_{C,k}^{t}\in\left\{0,1\right\}$}
                \STATE $f_{C,k}=f_{C,k}^{t}$.
                \STATE \textbf{break}
            \ELSE
                \STATE Record the average cooperator frequency over the last 1000 rounds as $f_{C,k}$.
            \ENDIF
        \ENDFOR
    \ENDFOR
    \STATE $f_{C}\left(r\right) =\left\langle f_{C,k}\right\rangle$.
\ENDFOR
\end{algorithmic}
\end{algorithm}

\section*{Data availability}
All data generated or analysed during this study are included within the paper and its supplementary information files.

\section*{Code availability}
All numerical calculations and computational simulations were performed in the C program (GCC version 8.1.0). All data analyses were performed in Matlab R2021b. All codes have been deposited into the publicly available repository at https://github.com/XCWang6/TempHyper.

\clearpage
\bibliographystyle{naturemag}
\bibliography{ref}

\clearpage

\makeatletter
\@fpsep\textheight
\makeatother

\clearpage

\clearpage

\clearpage

\begin{figure}
\centering
\includegraphics[width=0.55\linewidth]{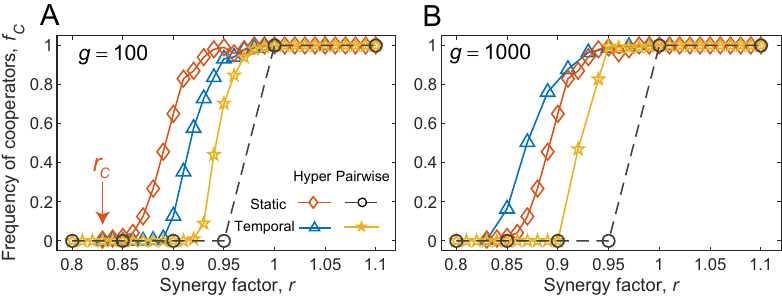}
\caption{\textbf{Advantages of higher-order interactions in promoting cooperation.}
We show the fraction of cooperation, $f_C$, as a function of the synergy factor, $r$, on empirical hypergraphs and on corresponding traditional networks with pairwise interactions, for $g=100$ (\textbf{a}) and $g=5000$ (\textbf{b}).
The value of $r_C$ is always smaller for temporal and static hypergraphs than that for traditional networks.
Other parameters are consistent with those of Fig.~\ref{fig:tem_den}.}
\label{fig:construction}
\end{figure}

\clearpage

\begin{figure}
\centering
\includegraphics[width=0.95\linewidth]{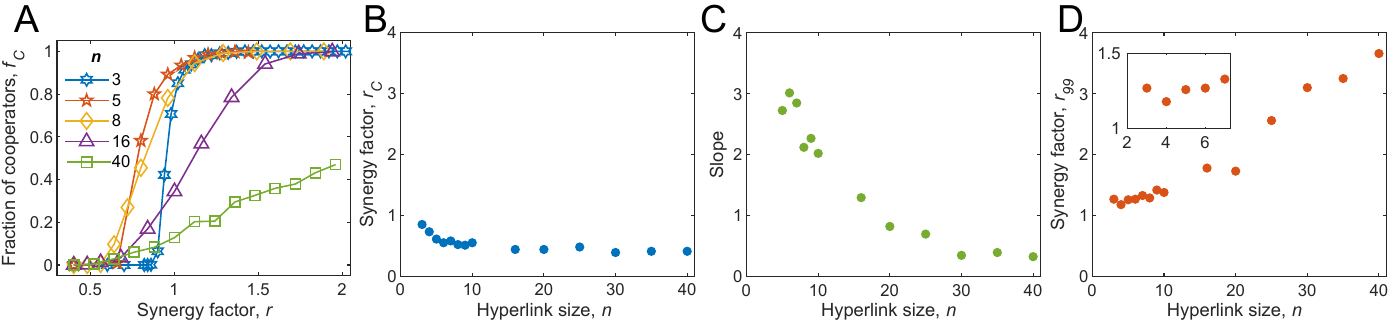}
\caption{\textbf{Moderate hyperedge sizes promote cooperation.}
\textbf{a}, We plot the fraction of cooperation, $f_C$, as a function of the synergy factor, $r$, for synthetic temporal hypergraphs with different hyperedge sizes, $n$.
As $n$ increases, cooperation is promoted at first and then impeded as it requires a larger value of $r$ to reach full cooperation.
We collect the values of $r_C$ (\textbf{b}) and $r_{99}$ (\textbf{d}) from simulations, calculate the fitted slopes when $0.2\le r\le0.8$ (\textbf{c}), and then plot their relationship with $n$. 
As $n$ increases, $r_C$ decreases gradually (\textbf{b}), while $r_{99}$ exhibits a slight initial decrease followed by a significant increase (\textbf{d}).
This corresponds to a drop in the fitted slopes with increasing $n$ (\textbf{c}).
For synthetic temporal hypergraphs, we use a population size of $N=100$ and the number of hyperedges on each subhypergraph to be $m=20$.
The number of interactions is $g=100$.
Other parameters are consistent with those used in Fig.~\ref{fig:tem_den}.
}

\label{fig: 5}
\end{figure}

\clearpage

\begin{figure}
\centering
\includegraphics[width=0.95\linewidth]{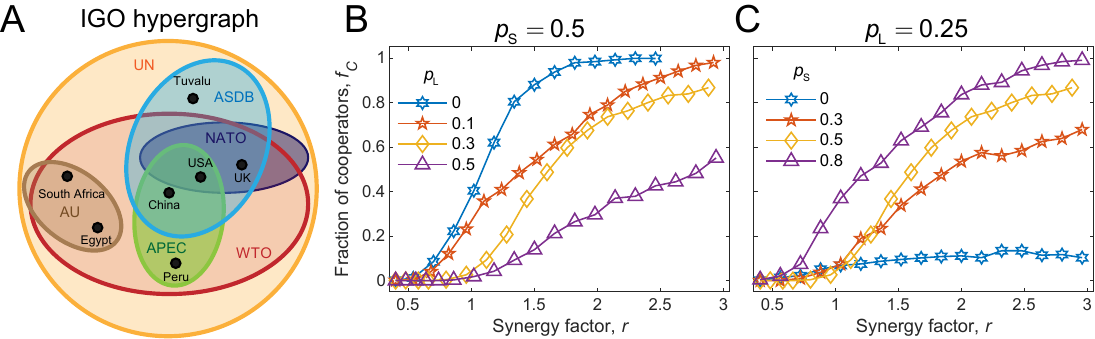}
\caption{
\textbf{Evolution of cooperation on intergovernmental organizations (IGO) temporal hypergraphs.}
\textbf{a},
An example of an IGO hypergraph. 
This hypergraph includes global organizations such as the United Nations (UN) and the World Trade Organization (WTO), as well as regional organizations like the Asia-Pacific Economic Cooperation (APEC), the Asian Development Bank (ASDB), the North Atlantic Treaty Organization (NATO), and the African Union (AU). 
In this hypergraph, each country is considered a node, and each organization is regarded as a hyperedge.
\textbf{b},
When the activation probability of small hyperedges (of size less than $40$) is held constant, increased activity of large hyperedges (of size at least $40$) tends to inhibit cooperation. 
\textbf{c},
Conversely, when the activation probability of large hyperedges remains constant, activating small hyperedges tends to promote cooperation.
See Methods for the construction of IGO temporal hypergraphs.
}
\label{fig: 6}
\end{figure}

\end{document}